\documentclass[twocolumn,twoside,slac_two]{revtex4}
\usepackage{graphicx}
\usepackage{fancyhdr}
\begin{document}

\title{Gauge-Invariant Localization of  Infinitely Many Gravitational Energies from All Possible Auxiliary Structures, Or, Why Pseudotensors Are Okay}

%

\author{J. Brian Pitts}
\affiliation{University of Notre Dame, IN 46556, USA}

\begin{abstract}

The problem of finding a  covariant expression for the distribution and conservation of gravitational energy-momentum dates to the 1910s.   A suitably covariant infinite-component localization is displayed, reflecting Bergmann's realization that there are infinitely many conserved gravitational energy-momenta.  Initially use is made of a flat background metric  or connection (or rather, all of them), because  the desired gauge invariance properties are obvious.  Partial gauge-fixing then yields an appropriate covariant quantity without any background metric or connection; one version is the  collection of pseudotensors of a given type, such as the Einstein pseudotensor, in \emph{every} coordinate system.  This solution to the gauge covariance problem is easily adapted to any pseudotensorial expression  or to any tensorial expression built with a background metric or connection.  Thus the specific functional form can be chosen on technical grounds such as relating to Noether's theorem and yielding expected values of conserved quantities in certain contexts and then rendered covariant using the procedure described here. The application to angular momentum localization is straightforward. Traditional objections to pseudotensors are based largely on the false assumption that there is only one gravitational energy rather than infinitely many. 

\end{abstract}

\maketitle

\thispagestyle{fancy}


\section{Introduction}

The problem of finding a  covariant expression for the distribution and conservation of gravitational energy-momentum for General Relativity dates to the 1910s.
Einstein took the requirement that the gravitational field equations alone entail energy-momentum conservation as a criterion for finding his field equations in his process of discovery (\cite{EinsteinEntwurfGerman,NortonField,Janssen,JanssenRenn}); ironically, it was widely concluded that the final  theory lacked any local conservation law for energy-momentum. The equation $\nabla_{\mu} T^{\mu\nu} =0$
for material stress-energy,
though a consequence of Einstein's equations, is a balance equation, not a conservation equation, because the covariant divergence of a rank 2 tensor (with any index placement and density weight) cannot be written using a coordinate divergence.  A coordinate divergence is required for integral conservation laws (\cite{Anderson}).  
 Gravitational energy-momentum has been reviewed on several occasions (\cite{TolmanEnergy,SchrodingerSTS,FletcherConservation,TrautmanConserve,CattaneoConsHP,CattaneoConsMilano,Davis,GoldbergReview,Carmeli,SzabadosReview}).  While there is no difficulty in writing down quantities satisfying local conservation laws (in the sense of a coordinate divergence), there seem to be too many  expressions without the anticipated interconnections.  More specifically, it has been expected that there ought to be a (10- or 16-component) tensor, geometric object, or other suitably covariant expression that describes the local distribution of gravitational energy-momentum, and yet evidently  there is not one.  Pseudotensorial answers go back to the Einstein's work in 1916 (\cite{EinsteinFoundationGerman}), while objections to them from Schr\"{o}dinger and from Bauer appeared in 1918 (\cite{SchrodingerEnergy,BauerEnergy,Pauli,Cattani}).  Later developments included  the introduction of additional background structures, such as a flat background metric (\cite{Rosen1,RosenAnn,Graiff,Bonazzola}), 
 an orthonormal tetrad (\cite{MollerAnnals}), or a flat connection (\cite{SorkinFlux,FatibeneFrancaviglia}).  While the introduction of such further structures has achieved  tensorial form with respect to coordinate transformations, this result has always come at the cost of introducing a new sort of  gauge dependence, because the choice of specific background metric, tetrad, or connection lacks physical meaning and yet affects the results.  The introduction of additional structures appears simply to move the lump in the carpet, not to flatten it out. Though new background structures 
 continue to be introduced, the inductive lesson only gets stronger that the gauge dependence problem is not resolvable in such a fashion (\cite{SzabadosReview}).  In this respect it is unclear that much has been gained beyond the original dependence of pseudotensors on coordinates found in the 1910s.

The solution to the problem of gauge dependence, briefly, is to take \emph{all possible auxiliary structures of a given type together}.  Thus, for example, the collection of all flat background metrics does not depend on the choice of any particular background metric.   Changing the flat background metric from one specific example to another merely leads to another member of the same collection. Looking for some finite-component expression that is covariant under a change of the background metric, though traditional, is a mistake.   Similar remarks hold for tetrads, connections, and even coordinate systems.  Indeed the cases of background metrics, background connections, and coordinate systems seem closely related, while the tetrad case differs and so will not be discussed much here. Its introduction of a gratuitous local Lorentz group is a major disadvantage, and it is in fact not required for spinors, as will appear below.  

Some authors, especially those who emphasize how different General Relativity is from other field theories rather than how similar it is, have tried to make the best out of the apparent non-existence of gauge-invariant gravitational energy localization.  Thus the question has been rejected as inappropriate, as shown by the equivalence principle:  
 ``[a]nybody who looks for a magic formula for `local gravitational energy-momentum' is looking for the right answer to the wrong question.'' (\cite[p. 467]{MTW})  However, this is an \emph{ad hoc} move.  Noether's theorems do not care about the equivalence principle; they simply give results in any coordinate system (\cite{BradingConserve}).  Rather than criticizing the results  of Noether's theorem in terms of preconceived notions of invariance and then mysteriously invoking a principle irrelevant to Noether's theorem  to reduce the puzzlement over the lack of an invariant energy complex, it is preferable to learn from the results of Noether's theorem that there is a broader notion of invariance suited to the existence of infinitely many distinct conserved energies.   There is no reason to expect the components of a pseudotensor to transform into each other once the vast multitude of gravitational energy-momenta is recognized.  Most issues discussed here are considered in more detail in a forthcoming paper (\cite{EnergyGravity}).


\section{Infinite-Component Covariant Density in Terms of All Flat Backgrounds}

Using a flat background metric tensor $\eta_{\mu\nu}$  allows one to describe gravitational energy in a tensorial way, independent of the choice of coordinates. Let $u$ represent bosonic matter fields; spinors will be considered below.  One can write down a gravitational energy-momentum tensor $t^{\mu\nu}[g_{\alpha\beta}, \eta_{\rho\sigma}]$ such that the total energy-momentum complex  $  (\sqrt{-g} T^{\mu\nu}[g_{\alpha\beta}, u]  +  \sqrt{-g} t^{\mu\nu}[g_{\alpha\beta}, \eta_{\rho\sigma}])$ satisfies covariant conservation
\begin{eqnarray}
 \partial_{\mu} (\sqrt{-g} T^{\mu\nu}  +  \sqrt{-g} t^{\mu\nu}) = 0
\end{eqnarray}
 with respect to the flat metric's torsion-free covariant derivative $ \partial_{\mu}.$   
 When General Relativity is formulated with a background metric, the action has two invariances, one under changes of coordinates and one under gauge transformations.  The latter transformations alter the mathematical relationship between $g_{\mu\nu}$ and $\eta_{\mu\nu}.$ For this reason $t^{\mu\nu}$ is tensorial with respect to coordinate transformations, but gauge-variant under gauge transformations  (\cite{Grishchuk}). 
Whereas  finite one-parameter  coordinate transformations can be written as 
\begin{eqnarray}
	g_{\sigma\rho} \rightarrow e^{\pounds_{\xi}} g_{\sigma\rho},
u \rightarrow e^{\pounds_{\xi}}   u,
	\eta_{\mu\nu} \rightarrow e^{\pounds_{\xi}} \eta_{\mu\nu}, \label{coord}
\end{eqnarray} 
 gauge transformations are written as 
\begin{eqnarray}
	g_{\sigma\rho} \rightarrow e^{\pounds_{\xi}} g_{\sigma\rho},
u \rightarrow e^{\pounds_{\xi}}   u,
	\eta_{\mu\nu} \rightarrow \eta_{\mu\nu}, \label{gauge}
\end{eqnarray} which leave the flat metric alone. 
Different and equally appropriate choices of background metric give different localizations, but correspond to the same physical situation.  Thus the achievement of tensorial energy-momentum localization has been purely formal; like a lump in the carpet, the  gauge dependence has merely been shifted, not ironed out.

One can avoid dependence on the choice of any particular background metric $\eta_{\mu\nu}$ by collecting all of them together in a set  $\{ (\forall \eta_{\rho\sigma}) \eta_{\rho\sigma} \}.$ Every flat metric yields a covariant conservation law: 
 \begin{eqnarray} \{ (\forall \eta_{\rho\sigma})  \;  \partial_{\mu}(T^{\mu\nu} \sqrt{-g} + t^{\mu\nu} \sqrt{-g}) =0    \}, \end{eqnarray}
 each  conserved using the appropriate flat covariant derivative.
Using a mere flat connection is analogous, but then there angular momentum problems (\cite{ChangNesterChen}, \emph{c.f.} \cite{GoldbergConservation}).  This is an infinite-component gauge-invariant localization of gravitational energy.  Gravitational energy is localized, but there are far more energies than one naively expected.  There is an apparently universal  tacit assumption that there  ought to be just one gravitational energy-momentum (with 10 or perhaps 16 components).  This assumption of uniqueness is especially clear in treatments by Goldberg
 (\cite{GoldbergReview}), Faddeev  (\cite{FaddeevEnergy}) and  Szabados  (\cite[section 3.1.3]{SzabadosReview}). Faddeev writes, ``The energy of the gravitational field is not localized, i.e., a uniquely defined energy density does not exist.'' (\cite{FaddeevEnergy}) While stated with special clarity in some cases, the assumption of uniqueness is implicit almost everywhere in the literature in the expectation that a pseudotensorial expression (perhaps Einstein's) in one coordinate system ought ideally to be related by a transformation law to that pseudotensor in another coordinate system in order to have the intended physical meaning of representing gravitational energy-momentum density.  
  This expectation of uniqueness makes sense if, as in other theories, there is only one energy in General Relativity.  It has been known at least since 1958 due to Bergmann and Komar, however, that there are \emph{infinitely many} gravitational energies, and that any coordinate basis or vector field generates one  (\cite{BergmannConservation,KomarConservation}).  Some of them might be zero; for example, a vector field derived by index-raising from an exact covector has vanishing Komar energy density. (The resulting Komar energies are unsatisfactory (\cite{PetrovKatz2}), so there is reason to expect the energies to depend on more than just a single vector field and the metric.) Some of the energies might plausibly regarded as faces of a single energy, such as if a Lorentz or affine transformation relates them.  But the point remains that there are a great many \emph{different} gravitational energy-momenta, uncountably infinitely many, far more than one naively expected. The question must be asked:  why can't they all be real? 
In fact there is no reason that they cannot all be real.  
Thus there is no reason whatsoever to expect distinct conserved quantities to behave mathematically as though they were just faces of one (finite-component) conserved quantity; the paradox dissolves. If there were a finite-component gauge-invariant localization, then it would represent under different gauges only different faces of the same entity.

One question not addressed here pertains to the uniqueness of the gravitational energy-momentum (pseudo)tensor, given the variety of candidates available.  It seems reasonable to require a candidate to be suitably related to Noether's theorem and to require correct values of integrated quantities in some basic contexts.  There might remain  some nonuniqueness due to the possibility of adding quantities with identically vanishing divergence.       A good candidate  is due to Joseph Katz,  Ji\v{r}\'{i} Bi\v{c}\'{a}k and Donald Lynden-Bell (\cite{KatzBicakLB,KatzEnergy,PetrovChapter}). Or perhaps the appropriate form depends on the boundary conditions (\cite{NesterQuasiPseudo,NesterQuasi}).


\section{Spinors as Almost Geometric Objects}

Given the most common ways of treating spinor fields, it is not obvious how gravitational energy localization in the form proposed here would work.  
M{\o}ller's orthonormal tetrad formalism was motivated in part by its supposed necessity to accommodate spinor fields (\cite{MollerAnnals}).   The  local Lorentz group  introduced in the tetrad formalism   seems quite unhelpful for localizing gravitational energy, however, even if one accepts all the tetrads at once.  Whereas the background metrics or background connections are closely related to the coordinate transformation freedom that is already present and ineliminable from the manifold, the local $O(3,1)$ group apparently bears no such relation. 

  Fortunately it is not the case that a tetrad is necessary for spinors, contrary to widely held opinion.    The tetrad formalism and local Lorentz group follow only if one insists on a linear coordinate transformation law for spinors as opposed to a nonlinear one (\cite[p. 234]{GatesGrisaruRocekSiegel} \cite{OPspinor}).  It is possible to include spinor fields almost like tensors in the Ogievetsky-Polubarinov-Bilyalov formalism (\cite{OPspinor,OP,BilyalovSpinors}). 
The spinor and  the metric  together form  a nonlinear geometric object $\langle g_{\mu\nu}, \psi  \rangle$  (\cite{OPspinor,OP,BilyalovSpinors}) (up to a sign for the spinor part), 
 with mild restrictions on the admissible coordinates  to distinguish the time coordinate from the spatial coordinates.  (The inequalities restricting the coordinates serve the same purpose as Bilyalov's matrix $T$ that interchanges two coordinates (\cite{BilyalovConservation}) to get time listed first. The possibility  of the field dependence of the admissible coordinates is typically not entertained when one defines a manifold as having all possible coordinate systems.)
 The nonlinearity is due to the fact that the new components of the spinor depend not only (linearly) on the old spinor components, but also on the metric  (\cite{OPspinor}).  By suitably weighting the spinor and exploiting conformal invariance, one could make the weighted spinor depend only on the conformal part of the metric.


\section{Localization in Terms of Pseudotensor in All Coordinate Systems}

 The use of a background metric or connection has the virtue that it manifestly has every sort of invariance that one would expect---both tensoriality under coordinate transformations and covariance under gauge transformations.  It is initially somewhat less clear what one should expect in a formalism with no background metric.  Fortunately one can gauge-fix the formalism above with a flat background metric or connection to find out.  I will ignore global issues by  pretending that all coordinate charts are defined everywhere.

One convenient  gauge fixing takes the bimetric formalism above and dispenses with the flat background metric tensors  by choosing  Cartesian coordinates for each flat metric separately.  Thus each flat metric tensor   $\eta_{\mu\nu}$ in the set $\{ (\forall \eta_{\rho\sigma})  \;  \eta_{\rho\sigma}  \}$  is downgraded to a \emph{matrix} $\eta_{MN}=diag(-1,1,1,1) $ and its resulting connection is downgraded to a three-index entity with only vanishing components, which can be ignored.  Now the former coordinate freedom   (\ref{coord})
 is destroyed, but the former gauge freedom (\ref{gauge})
 is formally converted into coordinate freedom (which has no effect on the numerical matrix $\eta_{MN}$).  The new coordinate freedom is still gauge freedom in the sense of Dirac-Bergmann constrained dynamics.  In a chart one has one's favorite pseudotensor 
$ t^{\mu\nu}[g_{\mu\nu}, \eta_{MN}],$ where the expression $g_{\mu\nu}$  now means the coordinate components of the curved metric. Using Einstein's field equations, the total energy-momentum complex  is conserved in the sense of having vanishing \emph{coordinate} divergence 
\begin{equation} \frac{\partial}{ \partial x^{\mu} }    (\sqrt{-g} T^{\mu\nu}  +  \sqrt{-g} t^{\mu\nu}[g_{\mu\nu}, \eta_{MN}])     =0 \end{equation}
in every coordinate system.  The gauge-invariant infinite-component gravitational energy-momentum distribution is just a certain pseudotensor in \emph{every} coordinate system.
The curved metric thus appears in all possible coordinate systems. 
This expression for the localization of gravitational energies has infinitely many components in a nontrivial sense:  each coordinate system picks out a distinct conserved energy.  The distinctness depends on the fact that the expression $t^{\mu\nu}$ is not a tensor or other geometric object (\cite{BergmannConservation,Anderson}). The components of a tensor or  any geometric object with respect to all coordinate systems give infinitely many faces of the same entity, but here we have infinitely many distinct entities, each appearing in its own adapted coordinate system.  

A long time ago Tolman proposed that having a pseudotensorial conservation law in every coordinate system is good enough, and forms an alternative way to be covariant  (\cite{TolmanEnergy,Tolman}).  He did not address the standard objections, however.  It is now clear that Tolman's proposal was correct as far as it went, but it needed to be supplemented with Bergmann's derivation of infinitely many different conservation laws from different coordinate bases.  

\section{Objections to Pseudotensors Wrongly Assume Uniqueness of Energy}

Having developed the covariant construction of localized energy-momenta, one can now easily resolve some standard objections to pseudotensors, which already appeared in Pauli's review (\cite{Pauli}) and have reappeared in countless places since then.  For example, it is noted with disappointment that a given pseudotensor (at least one without second derivatives) can be made to vanish at any point or along any worldline by a suitable choice of coordinates.  With the tacit assumption that gravitational energy-momentum is unique, one then concludes that there is no real fact of the matter pertaining to the density of gravitational energy-momentum at that point or along that worldline.  But the point or worldline was arbitrary, so there is no fact of the matter about gravitational energy-momentum localization in general.  Sometimes it is held that the situation  improves somewhat when symmetries yield Killing vectors, as in the case of spherical symmetry (\cite[p. 603]{MTW}.)  It is now clear how this objection goes astray:  the components of a given pseudotensor with respect to different coordinate systems in fact pick out \emph{different energies}, some but not all of which vanish at the arbitrarily chosen point or along the arbitrarily chosen worldline.  The fact that some energies vanish there but others don't is a bit unfamiliar, but it is in no way paradoxical on reflection.

Given long disappointment with gravitational energy localization, many authors have turned to seeking quasilocalization, in which the energy in some volume is specified, rather than the energy density at a point. Quasilocal energy is generally expected to be unique.  The injustice of that expectation, however, follows from the multitude of local energy densities pointed out by Bergmann (\cite{BergmannConservation}).   Pseudotensors are related to quasilocal methods (\cite{NesterQuasiPseudo,NesterQuasi}).   It is sometimes expected that a good quasilocal mass (energy) should vanish in flat spacetime, though that criterion does not hold for every proposed definition (\cite{Bergqvist}).  Likewise positive definiteness is  sometimes expected, though not always achieved (\cite{Bergqvist,SzabadosReview}). Local  gravitational energy-momentum expressions do not reliably vanish in Minkowski space-time for all gauges either; instead they vanish in some coordinate systems/gauges (\cite{PetrovChapter}) but not others.  If this result seems problematic, the resolution, again, is to notice that different coordinate systems/gauges pick out different energies.  It is a bit surprising that some of them fail to vanish even in Minkowski space-time, but it is not absurd.  Minkowski space-time is perhaps  unusual in that \emph{there exists} an energy-momentum density that vanishes everywhere.

Concerning Bauer's objection that flat spacetime in unimodular spherical coordinates has nonzero Einstein pseudotensor energy density (\cite{BauerEnergy,Pauli}), the fact that the same  pseudotensorial expression in  different coordinate systems picks out different energies removes the paradox. The fact that the total energy in these spherical coordinates diverges (\cite[p. 176]{Pauli}) is not terribly  surprising, given that spherical coordinates have marvelously strong coordinate effects.

Another traditional objection, this one due to Schr\"{o}dinger, calls attention to the vanishing of an Einstein pseudotensor (outside the Schwarzschild radius) for the Schwarzschild space-time in nearly Cartesian coordinates  with the unimodular condition $\sqrt{-g}=1$ (\cite{SchrodingerEnergy,Pauli}). 
Part of the worry presumably is that a vanishing Einstein pseudotensor suggests that no gravitational energy is present, but intuitively surely there is some present.  
 Once again the existence of many distinct energy densities is helpful to recognize.  Possibly one would expect the \emph{total} mass-energy to come out ``right''  in this context, but various localizations are known to exist, in some cases with the energy all in some small region, in others not (\cite{PetrovPoint,PetrovChapter}). If Schr\"{o}dinger had shown that \emph{all} the gravitational energy densities vanished outside the Schwarzschild radius, such a  result might be worrisome, but no such thing was shown.  That his particular energy vanishes is an interesting feature of gravitational energy as defined by the Einstein pseudotensor and his coordinate system, but it is no real objection. It is analogous to concluding that the electromagnetic field vanishes because one can choose a gauge with $A_0 = 0.$

\section{Equivalence of All Conservation Laws to Einstein's Equations}

In a typical field theory, one achieves energy-momentum conservation by noting that every field present in the equations of motion either has  Euler-Lagrange equations or has generalized Killing vector fields in the sense of vanishing Lie derivative (\cite{TrautmanUspekhi}).  In General Relativity as typically formulated (without a background metric or connection), every field present has Euler-Lagrange equations; there are no non-variational fields.  One might then expect that the energy-momentum of matter and gravity together to be conserved using both the gravitational field equations and the matter field equations.  A distinctive feature of General Relativity is that, because of gravitational gauge invariance (see, \emph{e.g.}, \cite{SliBimGRG}), conservation follows using the gravitational field equations alone, without using the matter equations (\cite{Anderson}). 
The collection of all of the pseudotensorial  conservation laws---a specific pseudotensor in all coordinates---is in fact \emph{equivalent} to Einstein's equations (\cite{Anderson,EnergyGravity}), so the reverse entailment also holds.
 This fact  sheds light on those approaches  that aim  to derive Einstein's field equations using the conservation laws as premises or lemmas (\cite{EinsteinEntwurfGerman,Deser,SliBimGRG}).

\section{Angular Momentum Localization}

 For angular momentum, one  introduces the coordinates $x^{\mu}$ and a symmetric choice of total energy-momentum complex $ \sqrt{-g} T^{\mu\nu}  +  \sqrt{-g} t^{\mu\nu}$  so that
\begin{equation} \mathcal{M}^{\mu\nu\alpha}\equiv   \sqrt{-g}( T^{\mu\nu}  +   t^{\mu\nu})  x^{\alpha} - \sqrt{-g}( T^{\mu\alpha}  +   t^{\mu\alpha})   x^{\nu}
\end{equation}
satisfies the conservation law $ \frac{ \partial}{\partial x^{\mu}} \mathcal{M}^{\mu\nu\alpha}=0 $ 
in all coordinates.  By parity of reasoning with the above, the collection of these angular momentum densities in \emph{every} coordinate system is an appropriate covariant infinite-component object.  Thus angular-momentum achieves a gauge-invariant localization in the same way as energy-momentum.

\section{Conceptual Benefits of Energy  Localization and Conservation}

If one is  aware of the uses to which the supposed lack of an energy conservation law in General Relativity has been put by now,  then the benefits of even a formal local energy conservation law become evident.  
The received view that there is no gauge-invariant and hence physically meaningful local conservation law for energy-momentum in General Relativity tends to inspire (though not strictly entail) a variety of unwarranted  conclusions. 
Some have criticized or rejected General Relativity (or Big Bang cosmology in particular) as having mystical tendencies on account of its supposed lack of conservation laws, while others have  appealed to General Relativity for certain purposes for the same reason.  
Elsewhere I discuss six such examples (\cite{EnergyGravity}).  The best known is due to Tryon, to the effect that the only meaningful energy conservation law for closed spaces is a global one with zero energy; thus it seems that energy conservation poses no objection to the spontaneous origin of universes   (\cite{Tryon}).
Finding gauge-invariant and hence physically meaningful local conservation laws  therefore contributes to scientific rationality by resolving a conceptual problem  (\cite{LaudanProgress}).

%
%
%
%
%


\bigskip 

\end{document}